\documentclass[twocolumn]{aastex63}
\UseRawInputEncoding

\newcommand{\jwst}{JWST}
\newcommand{\galname}{S04590}
\newcommand{\izw}{I\,Zw\,18}
\newcommand{\jykms}{Jy\,km\,s$^{-1}$}
\newcommand{\oii}{O\,{\sc ii}}
\newcommand{\oiii}{O\,{\sc iii}}
\newcommand{\cii}{C\,{\sc ii}}
\newcommand{\hi}{H\,{\sc i}}
\newcommand{\hii}{H\,{\sc ii}}

\usepackage{amsmath}
\usepackage{color}

\received{\today}
\revised{\today}
\accepted{\today}
\submitjournal{ApJ}

\shorttitle{The gas and stellar content of a metal-poor galaxy at $z=8.496$}
\shortauthors{Heintz et al.}

\begin{document}

\title{The gas and stellar content of a metal-poor galaxy at $z=8.496$ as revealed by JWST and ALMA}

\correspondingauthor{K.~E.~Heintz}
\email{keheintz@nbi.ku.dk}

\author[0000-0002-9389-7413]{K.~E.~Heintz}
\affiliation{Cosmic Dawn Center (DAWN), Denmark}
\affiliation{Niels Bohr Institute, University of Copenhagen, Jagtvej 128, DK-2200 Copenhagen N, Denmark}
\author[0000-0001-9419-9505]{C.~Gim\'enez-Arteaga}
\affiliation{Cosmic Dawn Center (DAWN), Denmark}
\affiliation{Niels Bohr Institute, University of Copenhagen, Jagtvej 128, DK-2200 Copenhagen N, Denmark}
\author[0000-0001-7201-5066]{S.~Fujimoto}
\affiliation{
Department of Astronomy, The University of Texas at Austin, Austin, TX 78712, USA
}
\affiliation{Cosmic Dawn Center (DAWN), Denmark}
\affiliation{Niels Bohr Institute, University of Copenhagen, Jagtvej 128, DK-2200 Copenhagen N, Denmark}
\author[0000-0003-2680-005X]{G.~Brammer}
\affiliation{Cosmic Dawn Center (DAWN), Denmark}
\affiliation{Niels Bohr Institute, University of Copenhagen, Jagtvej 128, DK-2200 Copenhagen N, Denmark}
\author[0000-0002-8726-7685]{D.~Espada}
\affiliation{Departamento de F\'{i}sica Te\'{o}rica y del Cosmos, Campus de Fuentenueva, Edificio Mecenas, Universidad de Granada, E-18071, Granada, Spain}
\affiliation{Instituto Carlos I de F\'{i}sica Te\'{o}rica y Computacional, Facultad de Ciencias, E-18071, Granada, Spain}
\author[0000-0001-9885-4589]{S.~Gillman}
\affiliation{Cosmic Dawn Center (DAWN), Denmark}
\affiliation{DTU-Space, Technical University of Denmark, Elektrovej 327, DK-2800, Kgs. Lyngby, Denmark}
\author[0000-0003-3926-1411]{J.~Gonz\'alez-L\'opez}
\affiliation{Las Campanas Observatory, Carnegie Institution of Washington, Casilla 601, La Serena, Chile}
\affiliation{N\'ucleo de Astronom\'ia, Facultad de Ingenier\'ia y Ciencias, Universidad Diego Portales, \\ Av. Ej\'ercito Libertador 441,  Santiago, Chile.}
\author[0000-0002-2554-1837]{T.~R.~Greve}
\affiliation{Cosmic Dawn Center (DAWN), Denmark} 
\affiliation{DTU-Space, Technical University of Denmark, Elektrovej 327, DK-2800, Kgs. Lyngby, Denmark}
\author[0000-0002-6047-430X]{Y.~Harikane}
\affiliation{Institute for Cosmic Ray Research, The University of Tokyo, 5-1-5 Kashiwanoha, Kashiwa, Chiba 277-8582, Japan}
\author[0000-0001-6469-8725]{B.~Hatsukade}
\affiliation{Institute of Astronomy, Graduate School of Science, The University of Tokyo, 2-21-1 Osawa, Mitaka, Tokyo 181-0015, Japan}
\author[0000-0002-7821-8873]{K.~K.~Knudsen}
\affiliation{Department of Space, Earth and Environment, Chalmers University of Technology, Onsala Space Observatory, SE-439 92 Onsala, Sweden}
\author[0000-0002-6610-2048]{A.~M.~Koekemoer}
\affiliation{Space Telescope Science Institute, 3700 San Martin Dr., Baltimore, MD 21218, USA}
\author[0000-0002-4052-2394]{K.~Kohno}
\affiliation{Institute of Astronomy, Graduate School of Science, The University of Tokyo, 2-21-1 Osawa, Mitaka, Tokyo 181-0015, Japan}
\affiliation{Research Center for the Early Universe, Graduate School of Science, The University of Tokyo, 7-3-1 Hongo, Bunkyo-ku, Tokyo 113-0033, Japan}
\author[0000-0002-5588-9156]{V.~Kokorev}
\affiliation{Kapteyn Astronomical Institute, University of Groningen, PO Box 800, 9700 AV Groningen, The Netherlands}
\author[0000-0002-2419-3068]{M.~M.~Lee}
\affiliation{Cosmic Dawn Center (DAWN), Denmark} 
\affiliation{DTU-Space, Technical University of Denmark, Elektrovej 327, DK-2800, Kgs. Lyngby, Denmark}
\author[0000-0002-4872-2294]{G.~E.~Magdis}
\affiliation{Cosmic Dawn Center (DAWN), Denmark}
\affiliation{DTU-Space, Technical University of Denmark, Elektrovej 327, DK-2800, Kgs. Lyngby, Denmark}
\affiliation{Niels Bohr Institute, University of Copenhagen, Jagtvej 128, DK-2200 Copenhagen N, Denmark}
\author[0000-0002-7524-374X]{E.~J.~Nelson}
\affiliation{Department for Astrophysical and Planetary Science, University of Colorado, Boulder, CO 80309, USA}
\author[0000-0001-9705-2461]{F.~Rizzo}
\affiliation{Cosmic Dawn Center (DAWN), Denmark}
\affiliation{Niels Bohr Institute, University of Copenhagen, Jagtvej 128, DK-2200 Copenhagen N, Denmark}
\author{R.~L.~Sanders}
\affiliation{Department of Physics and Astronomy, University of California, Los Angeles, 430 Portola Plaza, Los Angeles, CA 90095, USA}
\affiliation{Department of Physics, University of California, Davis, 1 Shields Avenue, Davis, CA 95616, USA}
\author[0000-0001-7144-7182]{D.~Schaerer}
\affiliation{Observatoire de Gen\`eve, Universit\'e de Gen\`eve, Chemin Pegasi 51, 1290 Versoix, Switzerland}
\affiliation{CNRS, IRAP, 14 Avenue E. Belin, 31400 Toulouse, France}
\author[0000-0003-3509-4855]{A.~E.~Shapley}
\affiliation{Department of Physics and Astronomy, University of California, Los Angeles, 430 Portola Plaza, Los Angeles, CA 90095, USA}
\author[0000-0002-6338-7295]{V.~B.~Strait}
\affiliation{Cosmic Dawn Center (DAWN), Denmark}
\affiliation{Niels Bohr Institute, University of Copenhagen, Jagtvej 128, DK-2200 Copenhagen N, Denmark}
\author[0000-0003-3631-7176]{S.~Toft}
\affiliation{Cosmic Dawn Center (DAWN), Denmark}
\affiliation{Niels Bohr Institute, University of Copenhagen, Jagtvej 128, DK-2200 Copenhagen N, Denmark}
\author[0000-0001-6477-4011]{F.~Valentino}
\affiliation{Cosmic Dawn Center (DAWN), Denmark}
\affiliation{Niels Bohr Institute, University of Copenhagen, Jagtvej 128, DK-2200 Copenhagen N, Denmark}
\affiliation{European Southern Observatory, Karl-Schwarzschild-Str. 2, D-85748 Garching bei Munchen, Germany}
\author[0000-0002-5027-0135]{A.~van der Wel}
\affil{Sterrenkundig Observatorium, Universiteit Gent, Krijgslaan 281 S9, 9000 Gent, Belgium}
\author[0000-0002-1905-4194]{A.~P.~Vijayan}
\affiliation{Cosmic Dawn Center (DAWN), Denmark}
\affiliation{DTU-Space, Technical University of Denmark, Elektrovej 327, DK-2800, Kgs. Lyngby, Denmark}
\author[0000-0002-4465-8264]{D.~Watson}
\affiliation{Cosmic Dawn Center (DAWN), Denmark}
\affiliation{Niels Bohr Institute, University of Copenhagen, Jagtvej 128, DK-2200 Copenhagen N, Denmark}
\author[0000-0002-8686-8737]{F.~E.~Bauer}
\affiliation{Instituto de Astrof{\'{\i}}sica, Facultad de F{\'{i}}sica, Pontificia Universidad Cat{\'{o}}lica de Chile, Campus San Joaquín, Av. Vicuña Mackenna 4860, Macul Santiago, Chile, 7820436} 
\affiliation{Centro de Astroingenier{\'{\i}}a, Facultad de F{\'{i}}sica, Pontificia Universidad Cat{\'{o}}lica de Chile, Campus San Joaquín, Av. Vicuña Mackenna 4860, Macul Santiago, Chile, 7820436} 
\affiliation{Millennium Institute of Astrophysics, Nuncio Monse{\~{n}}or S{\'{o}}tero Sanz 100, Of 104, Providencia, Santiago, Chile} 
\author{C.~R.~Christiansen}
\affiliation{Cosmic Dawn Center (DAWN), Denmark}
\affiliation{Niels Bohr Institute, University of Copenhagen, Jagtvej 128, DK-2200 Copenhagen N, Denmark}
\author{S.~N.~Wilson}
\affiliation{Cosmic Dawn Center (DAWN), Denmark}
\affiliation{Niels Bohr Institute, University of Copenhagen, Jagtvej 128, DK-2200 Copenhagen N, Denmark}

\begin{abstract}
We present a joint analysis of the galaxy \galname\ at $z=8.496$ based on NIRSpec, NIRCam, and NIRISS observations obtained as part of the Early Release Observations program of the James Webb Space Telescope (\jwst) and the far-infrared [\cii]-$158\mu$m emission line detected by dedicated Atacama Large Millimeter/submillimeter Array (ALMA) observations.
We determine the physical properties of \galname\ from modeling of the spectral energy distribution (SED) and through the redshifted optical nebular emission lines detected with \jwst/NIRSpec. The best-fit SED model reveals a low-mass ($M_\star = 10^{7.2}-10^{8}\,M_{\odot}$) galaxy with a low oxygen abundance of $12+\log{\rm (O/H)} = 7.16^{+0.10}_{-0.12}$ derived from the strong nebular and auroral emission lines. 
Assuming that [\cii] effectively traces the interstellar medium (ISM), we estimate the total gas mass of the galaxy to be $M_{\rm gas} = (8.0\pm 4.0)\times 10^{8}\,M_\odot$ based on the luminosity and spatial extent of [\cii]. 
This yields an exceptionally high gas fraction, $f_{\rm gas} = M_{\rm gas}/(M_{\rm gas} + M_\star) \gtrsim 90\%$, though one still consistent with the range expected for its low metallicity. 
We further derive the metal mass of the galaxy based on the gas mass and gas-phase metallicity, which we find to be consistent with the expected metal production from Type II supernovae. Finally, we make the first constraints on the dust-to-gas (DTG) and dust-to-metal (DTM) ratios of galaxies in the epoch of reionization at $z\gtrsim 6$, showing overall low mass ratios of logDGT $<-3.8$ and logDTM $<-0.5$, though consistent with established scaling relations and in particular the local metal-poor galaxy I Zwicky 18. Our analysis highlights the synergy between ALMA and \jwst\ in characterizing the gas, metal, and stellar content of the first generation of galaxies.
\end{abstract}

\section{Introduction} \label{sec:intro}

The first epoch of galaxy formation is believed to have occurred at $z\sim 15-20$ \citep{Hashimoto18,Robertson22}, some 100-200 million years after the Big Bang. This process was initiated by the infall of pristine, neutral gas onto the first protogalactic halos, which would eventually cool and condense into H$_2$ gas reservoirs and later form stars \citep{Dayal18}. Ending their lives as core-collapse supernovae, the first generation of stars enriched the pristine gas with dust and metals, providing the material for subsequent star formation and galaxy evolution, heralding the epoch of reionization. 
It is therefore vital to probe the interstellar medium (ISM) and dust and metal build-up in galaxies at $z\gtrsim 6$ to constrain the star-formation histories (SFHs) of the first galaxies. 

The Atacama Large Millimeter/sub-millimeter Array (ALMA) has been paramount in providing a clear census of the rest-frame far-infrared (FIR) emission of galaxies at $z\gtrsim 6$. The fine-structure $^2P_{3/2}-^2P_{1/2}$ [\cii]$-158\mu$m line transition in particular, which is one of the major ISM cooling lines \citep{Hollenbach99,Wolfire03,Carilli13,Lagache18}, has been used to infer the ISM dynamics \citep{Capak15,Smit18,Matthee19} and overall gas content \citep[][M. Aravena et al., in prep.]{Knudsen16,DessaugesZavadsky20,Heintz21,Heintz22} of individual galaxies at $z\gtrsim 6$. In more local galaxies at $z\approx 0$, [\cii] has been linked to the star-formation rate (SFR) \citep{Stacey91,DeLooze14,Cormier15} and physically associated with the neutral, atomic gas and photodissociation regions (PDRs) \citep{Madden93,Madden97,Hollenbach99}. The relationship between the SFR and the luminosity of [\cii], $L_{\rm [CII]}$, has been shown to be universally valid out to $z\approx 8$ \citep{Schaerer20}. Further, the [\cii] halos surrounding these early galaxies, which typically extend $2-3\times$ beyond their UV components \citep[e.g.,][]{Smit18,Fujimoto19,Matthee19,Carniani20,Fudamoto22}, are reminiscent of the extended [\cii] emission originating from the neutral, atomic hydrogen (\hi) gas phase in local galaxies \citep{Madden93}. FIR continuum emission \citep{Draine07,Scoville14} has further enabled strong constraints on the dust mass and temperature of galaxies at $z\gtrsim 6$ \citep{Bakx21,Ferrera22,Dayal22,Sommovigo21,Sommovigo22,Inami22,Algera23}. Thus, we now have a robust census of the dust and gas properties of galaxies well into the epoch of reionization.    

The final missing ingredient to understand and fully characterize the ISM is the metal abundance and the overall chemical enrichment of the galaxies at this early epoch. Strong-line diagnostics of the most prominent nebular emission lines have been used as viable tools to infer the gas-phase metallicities \citep{Kewley08,Kewley19,Maiolino19}, typically calibrated to the more direct $T_e$ method. However, these measurements have previously been limited by the atmospheric cutoff for ground-based observations and have as a result mostly been possible to derive robustly only for galaxies up to $z\approx 3$ \citep{Christensen12,Sanders21}. Chemical abundances have also been derived for foreground galaxies toward background sources such as gamma-ray bursts and quasars out to $z\gtrsim 6$ \citep{Hartoog15,Simcoe20,Saccardi22}, which, however, only represent a single line of sight through the absorbing galaxies. 

Now with the advent of \jwst\ we are able to derive robust, total gas-phase metallicities through the strong-line diagnostics of the most prominent nebular emission lines of galaxies during the first Gyr of cosmic time \citep[e.g.,][]{Rhoads22,Schaerer22,Curti23,Tacchella22,Heintz22b}. Crucially, these observations have also revealed prominent detections of the auroral [\oiii]\,$\lambda 4363$ emission line which enables direct $T_e$-based metallicity measurements, necessary to consider since the typical strong-line diagnostics have yet only been calibrated to local $z\approx 0$ galaxy samples. \jwst\ and ALMA thus jointly provide a unique and comprehensive view of the gas, dust, and metal content of the first galaxies. While the first such combined studies of some of the highest-redshift galaxy candidates identified by \jwst\ remain inconclusive \citep{Yoon22,Bakx22,Popping22b,Fujimoto22}, it is still not entirely clear whether this is due to their likely low metallicities or to their potentially erroneous photometric redshifts.

Here we present a characterization of the stellar, gas, and metal content of a galaxy at $z\approx 8.5$, from a joint analysis of observations obtained with JWST and ALMA. 
Hereafter, we refer to this galaxy as \galname, based on its ID in the JWST/NIRSPEC mask design.
The \jwst\ data, including observations taken with the Near-infrared Spectrograph (NIRSpec), the Near-infrared camera (NIRCam), and the Near Infrared Imager and Slitless Spectrograph (NIRISS), are from the early-release science observations (ERO) covering the SMACS J0723.3–7327 galaxy cluster (hereafter SMACS0723), \citep[see][]{Pontoppidan22}. The ALMA observations were presented in detail by \citet{Fujimoto23}, who reported the detections of the FIR emission lines [\oiii]-$88\mu$m and [\cii]-$158\mu$m of \galname. This galaxy thus also presents the most distant detection of the [\cii]-$158\mu$m line transition feature to date, and greatly highlights the importance of having robust, spectroscopic redshifts from JWST to optimize the follow-up at FIR wavelengths with ALMA.

We structure the paper as follows. In Section~\ref{sec:obs} the observations are detailed and in Section~\ref{sec:res} we present and analyze the results. In Section~\ref{sec:disc} we infer the gas and metal content of the target galaxy and discuss the implications of this work. In Section~\ref{sec:conc} we highlight the great synergy between \jwst\ and ALMA, and provide some concluding remarks. Throughout the paper we quote magnitudes in the AB system and adopt the concordance $\Lambda$CDM cosmological model with $\Omega_{\rm m} = 0.315$, $\Omega_{\Lambda} = 0.685$, and $H_0 = 67.4$\,km\,s$^{-1}$\,Mpc$^{-1}$ \citep{Planck18}. We assume a \citet{Chabrier03} initial mass function (IMF) and the solar abundances from \citet{Asplund09} with $Z_\odot = 0.0134$. 


\section{Observations} \label{sec:obs}

\subsection{NIRCam and NIRISS photometry}

We include the \jwst\ NIRCam and NIRISS photometric data of \galname\ from the catalog presented by G. Brammer et al. (in preparation), which provides a compilation of the JWST ERO photometric data released to date. Briefly, the raw data entered into this catalog has been reduced using the public software package \texttt{grizli} \citep{Brammer19}, which masks imaging artifacts, provides astrometric calibrations based on the Gaia DR3 catalog, and shifts the images to a common pixel scale of 0\farcs04 pixel$^{-1}$ using \texttt{astrodrizzle}. Cutouts of the reduced images from NIRCam and NIRISS used in this analysis are provided on a dedicated repository\footnote{\url{https://github.com/keheintz/S04590}}. \galname\ is detected in the five deep NIRCam imaging filters F150W, F200W, F277W, F356W, and F444W, with an F150W ($\approx H-$band) magnitude = 27.3. The non-detection in the bluest filter F090W makes this galaxy a clear photometric dropout. We also include two shallower images obtained with NIRISS with the F115W and F200W filters. This set of photometric measurements is derived using the updated zero-points, and corrected for Milky Way extinction. Since the source is extended in the images, we use the aperture photometry derived using a diameter of 0\farcs5 from the catalog by G. Brammer et al. (in preparation). 

\begin{figure*}[t!]
\centering
\includegraphics[width=17cm]{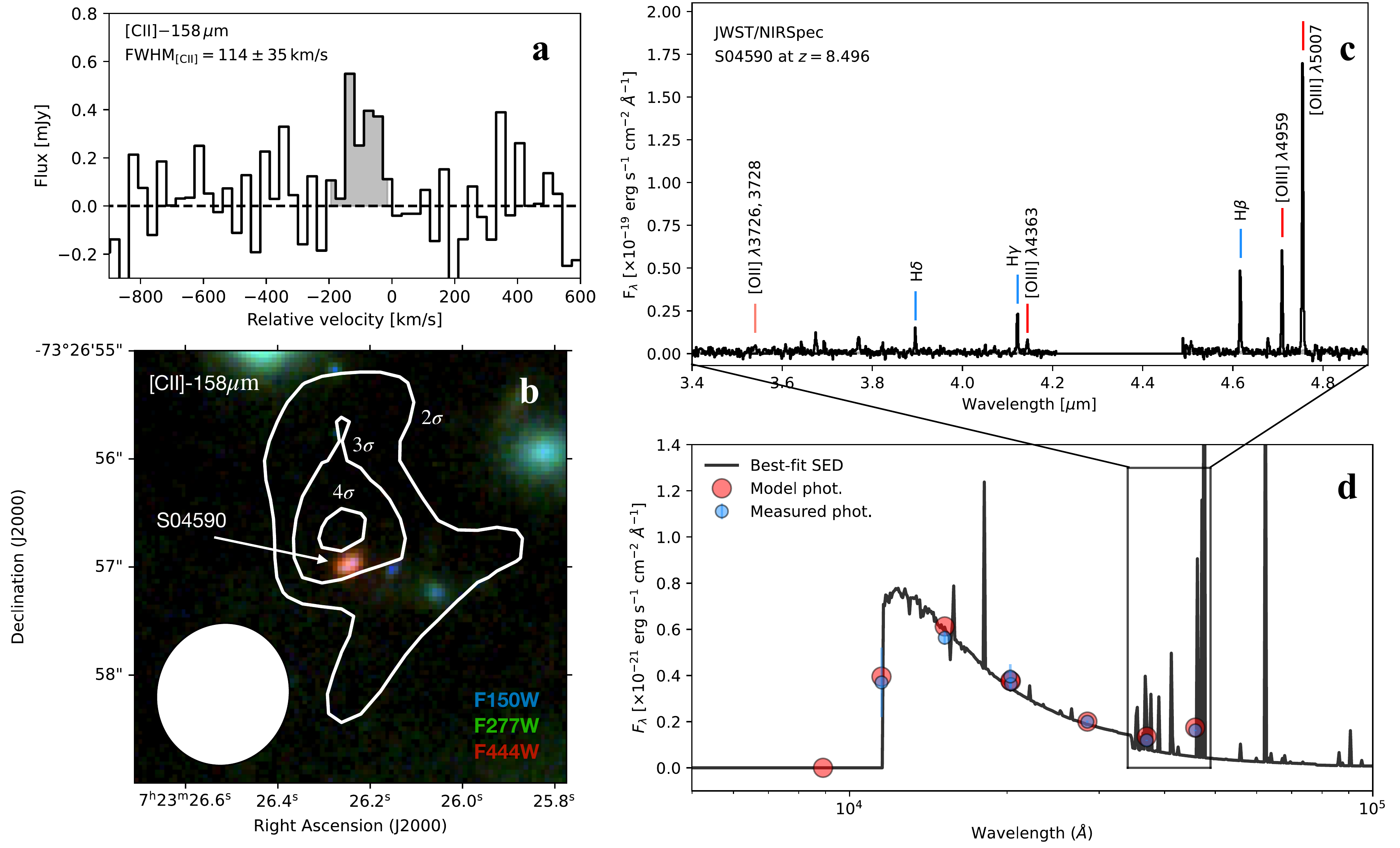}
\caption{Panel (a): [\cii] line spectrum from the 30 km\,s$^{-1}$ ALMA data cube \citep{Fujimoto23}. Panel (b): False-color RGB \jwst/NIRCam image zoomed in on \galname (blue: F150W; green: F277W; red: F444W), overlaid with the $2\sigma$, $3\sigma$, and $4\sigma$ contours of the [\cii] moment-0 map with the beam size represented by the white ellipse. Panel (c): \jwst/NIRSpec spectrum with the most prominent nebular and auroral emission lines marked. Panel (d): Best-fit SED model (black curve) and model photometry (red points) based on the derived \jwst/NIRCam+NIRISS photometry (blue points) using {\sc Bagpipes} \citep{bagpipes} and assuming the dust attenuation curve from \citet{Salim18}.}
\label{fig:sedspec}
\end{figure*}

The centroid of the galaxy is localized to R.A., Decl. = $07^{\rm h}23^{\rm m}26.24^{\rm s}$, $-73{^\circ}26^{\prime}57\farcs0$.
A zoomed-in image showing the false RGB color (F150W+F277W+F444W) of \galname\ is shown in Fig.~\ref{fig:sedspec}, together with the spectral energy distribution (SED) inferred from modeling of the photometric data points (see also Sect.~\ref{ssec:sed}). 

\subsection{NIRSpec spectroscopy}

The \jwst\ ERO spectroscopic data of \galname\ have already been presented and examined by a number of independent studies \citep[][]{Carnall23,Trump22,Rhoads22,Schaerer22,Katz23,Curti23,ArellaniCordova22,Taylor22,Brinchmann22,Tacchella22}. The NIRSpec spectroscopy is obtained with the medium-resolution gratings G235M/F170LP ($1.75-3.1\mu$m) and G395M/F290LP ($2.9-5.2\mu$m), with $\mathcal{R} = \lambda/\Delta \lambda = 1000$ \citep{Jakobsen22}, for a total exposure time of 8754\,s. We use the JWST pipeline to perform the standard wavelength, flat-field, and photometric calibrations to the individual NIRSpec exposure files\footnote{i.e., calibration levels 1 and 2 with \texttt{jwst} version 1.8.2}. We generate the full combined 2D spectra and optimal \citep{Horne86} 1D extractions with scripts that extend the standard pipeline functionality \citep{msaexp}. We further scale the overall flux density of the spectrum to the derived photometry to improve the absolute flux calibration by matching the integrated flux within the F444W passband to the derived photometry of the same filter. The reduced spectrum is made available on the dedicated repository$^1$.
Overall, our extracted version of the spectrum appears consistent with the publicly available 1D spectrum from \citet{Curti23}, though we detect a slightly weaker [\oii]\,$\lambda\lambda 3726,3729$ doublet and [\oiii]\,$\lambda 4363$ line transition (see Sect.~\ref{ssec:spec} for further discussion). We confirm a similar trend in another up-to-date NIRSpec study by \citet{Nakajima23}. Both of these spectra were derived from an independent source extraction and calibration, which alleviated the issues regarding the flux calibration and the inferred line ratios from the Phase 3 MAST products due to an error in the first s007 exposure. 
The spectrum is presented in Fig.~\ref{fig:sedspec}, where the most prominent nebular emission lines used to determine the redshift and infer the physical properties of the galaxy are marked as well.

\subsection{ALMA observations}

ALMA Band~5 follow-up observations of [\cii] spectroscopy were performed as a part of the Cycle~8 DDT program (\#2022.A.00022.S, PI: S. Fujimoto).  
The details of the observations, data reduction, calibrations, and imaging procedures are presented in \citet{Fujimoto23}. 
Briefly, we reduced and calibrated the ALMA data with the Common Astronomy Software Applications package version 6.4.1.12 (CASA; \citealt{casa2022}) by using the pipeline script in the standard manner. 
We adopted the natural weighting, resulting in an FWHM size of the synthesized beam of $1\farcs35\times1\farcs25$ with $1\sigma$ sensitivities for the continuum map and the line cube with a 40~km~s$^{-1}$ channel width of 11.6 $\mu$Jy~beam$^{-1}$ and 165 $\mu$Jy~beam$^{-1}$, respectively. 

The [\cii] line emission originating from \galname\ is detected at $4.5\sigma$, but is blueshifted by 90\,km\,s$^{-1}$ from the rest-frame optical spectrum and with a spatial offset of $0\farcs5$ from the rest-frame UV component \citep[][see also Fig.~\ref{fig:sedspec}]{Fujimoto23}. The robustness of the detection is substantiated by the simultaneous detection of the FIR [\oiii]-$88\mu$m emission line, which, however, is spatially and spectroscopically consistent with the UV, star-forming component. 

The [\cii] line is detected with a velocity-integrated flux of $S_{\rm [CII]} = 0.094\pm 0.027$\,\jykms. This corresponds to an intrinsic luminosity \citep[following, e.g.,][]{Solomon05} of $L_{\rm [CII]} = (1.67\pm 0.37)\times 10^{7}\,L_\odot$, corrected for the magnification factor due to gravitational lensing, which we here assume to be $\mu = 8.69$ from the updated {\sc Glafic} lens model \citep{Harikane22}. We note the 10-30\% differences among the most recent mass models for this cluster \citep{Caminha22,Mahler22}. To represent this uncertainty and for internal consistency with \citet{Fujimoto23} we adopt $\mu = 8.69\pm 2.61$ here.
Modeling the [\cii] emission line in the 1D spectrum extracted from the higher-resoluation data cube, with a resolution of $30$\,km\,s$^{-1}$, yields an FWHM of ${\rm FWHM}_{\rm [CII],obs} = 118\pm 36\,$km\,s$^{-1}$ \citep{Fujimoto23}, or a line width of $\sigma_{\rm [CII],obs} = 50\pm 15\,$km\,s$^{-1}$. Correcting for the instrumental resolution this corresponds to an intrinsic line width of ${\rm FWHM}_{\rm [CII],intr} = \sqrt{{\rm FWHM}_{\rm [CII],obs}^2 - {\rm FWHM}_{\rm inst}^2} = 114\pm 35\,$km\,s$^{-1}$, as reported in Table~\ref{tab:galprops} and used throughout.

\section{Analysis \& Results} \label{sec:res}

\begin{deluxetable}{lr}
  \tabletypesize{\normalsize}
  \tablecolumns{2}
  \tablecaption{Measurements and physical properties of \galname.\label{tab:galprops}}
  \smallskip
  \tablehead{}
   \startdata
    %
    R.A. (J2000) & 07:23:26.24 (110.85933 deg) \\
    Decl. (J2000) & $-$73:26:57.0 ($-$73.44917 deg) \\
    \rule{0pt}{4ex}$z_{\rm spec}$& $8.4959\pm 0.0003$ \\  
    \tablenotemark{a}$\mathrm{log}(M_\star/M_\odot)$ & $7.15\pm0.15$\\
    \tablenotemark{a}Mean stellar age (Myr) &  $1.3^{+0.4}_{-0.3}$\\
    \tablenotemark{a}$A_V$ (mag) & $0.20^{+0.04}_{-0.03}$ \\
    \tablenotemark{a}$\log_{10}(U)$ & $-1.12^{+0.08}_{-0.18}$\\
    \tablenotemark{a}$\beta_{\rm UV} $ & $-1.81\pm 0.08$\\
    \rule{0pt}{4ex}\tablenotemark{b}$S_{\rm [CII]}$ (Jy\,km\,s$^{-1}$)& $(1.1\pm 0.3)\times 10^{-2}$ \\ 
    \tablenotemark{b}$L_{\rm [CII]}$ ($L_\odot$) & $(1.67\pm 0.37)\times10^{7}$ \\  
    $\rm \sigma_{\rm [CII]}$ ($\mathrm{km\,s^{-1}}$)& $50\pm15$\\
    $\rm FWHM_{\rm [CII]}$ ($\mathrm{km\,s^{-1}}$)& $114\pm35 $\\
    \rule{0pt}{4ex}SFR$_{\rm H\beta}$ ($M_\odot \, {\rm yr}^{-1}$) & $2.9\pm 0.9$ \\
    $T_e$ (K) & $(2.4\pm 0.4) \times 10^{4}$ \\
    $12+\log{\rm (O/H)}$ & $7.16^{+0.10}_{-0.12}$ \\
    $\log Z/Z_\odot$ & $-1.53^{+0.10}_{-0.12}$ \\
    \rule{0pt}{4ex}$M_{\rm dyn,tot}$ ($M_\odot$) & $(9.0\pm 4.5)\times 10^{8}$ \\
    $M_{\rm gas}$ ($M_\odot$) & $(8.0\pm 4.0)\times 10^{8}$ \\
    $M_{\rm Z,ISM}$ ($M_\odot$) & $(3.2\pm 1.6)\times 10^{5}$ \\
    \tablenotemark{b}$M_{\rm d}$ ($M_\odot$) & $< 1.2\times 10^{5}$ \\
     \enddata
  \tablecomments{All values reported here that are affected by magnification due to gravitational lensing have been corrected by the magnification factor $\mu = 8.69$.} \tablenotetext{a}{Inferred physical properties from the SED model are based on the integrated photometry (see text for caveats).}
  \tablenotetext{b}{Measurements from \citet{Fujimoto23}.}
\end{deluxetable}

\subsection{Morphology and galaxy size}\label{ssec:morph}

We first model the morphology of \galname\ in all the NIRCam images in which it is detected (excluding F090W) simultaneously using \textsc{Galfit} \citep{Peng02}. We find that a simple Sers\'ic profile is able to reproduce the observed light profile, with an effective radius around $1.5-2$ pixels depending on the filter used. In the F150W filter (rest-frame UV $\approx 1600\,\AA$) we derive an effective radius of $r_{\rm e,UV} = 0\farcs07$, corresponding to $r_{\rm e,UV} = 0.33$\,kpc in the image plane at $z=8.496$ \citep[consistent with the results of][]{Rhoads22}. Correcting for the magnification factor, we estimate a physical size of $r_{\rm e,UV} / \sqrt{\mu} = 0.11$\,kpc in the source plane.

\subsection{Modeling the spectral energy distribution}\label{ssec:sed}

To infer the physical properties as well as constrain the SFH of \galname, we model the integrated SED using the fitting code \textsc{Bagpipes} \citep[][]{bagpipes}. We fix the redshift to the spectroscopic NIRSpec redshift $z_{\textrm{spec}}=8.496$ (see Sect.~\ref{ssec:spec}). We include nebular emission via \textsc{Cloudy} \citep[][]{cloudy}, allowing the ionization parameter, $U$, to vary within $-3 < \log_{10}(U) < -1$ to account for the typically higher ionization parameter at $z\gtrsim 6$ \citep{Sugahara22}, and the derived electron temperature \citep[see, e.g.,][and the section below]{Trump22,Curti23,ArellaniCordova22,Katz23}. We assume a \citet{Salim18} attenuation curve, which is notably steeper than the typically adopted ``Calzetti'' extinction curve \citep{Calzetti00}. The steeper \citet{Salim18} curve better represents low-mass, low-metallicity galaxies \citep{Reddy18} and effectively outputs a lower $A_V$. We use a constant SFH model, which mimics a bursty SFH on short timescales, expected for young galaxies at this redshift. A more detailed description of the priors used in the modeling, together with the resulting posterior distributions, is provided in Appendix A. 

The best-fit SED model based on the integrated photometry shown in Fig.~\ref{fig:sedspec} yields a stellar mass $\log_{10}(M_\star/M_{\odot}) = 7.15\pm0.15$, a young stellar population with a mean stellar age of $1.3^{+0.4}_{-0.3}$\,Myr, a relatively blue UV continuum slope $\beta = -1.81\pm 0.08$, and a low dust attenuation $A_V = 0.20^{+0.04}_{-0.03}$\,mag for \galname, as summarized in Table~\ref{tab:galprops}. This is overall consistent with the results from \citet{Carnall23}. We note that \citet{GimenezArteaga23} found a substantially higher stellar mass of $M_\star \approx 10^{8}\,M_\odot$ based on a resolved SED study of a set of galaxies at $z\gtrsim 5$ (including \galname), which has also been recovered assuming a burstier SFH \citep{Tacchella22}. To take into account any potential systematic uncertainties on the stellar mass estimates from varying SFH models \citep{Whitler23} or in general across SED fitting codes \citep{Pacifici22}, we assume a more conservative stellar mass of $M_\star = 10^{7.2}-10^{8.0}\,M_\odot$ in the following analysis.

\subsection{Rest-frame nebular line emission and metallicity} \label{ssec:spec}

Based on the extracted 1D NIRSpec spectrum, we derive the line fluxes of the following nebular and auroral emission lines: [\oiii]\,$\lambda\lambda\lambda 4363, 4959, 5007$; the doublet [\oii]\,$\lambda\lambda 3726,3729$; and the Balmer lines H$\delta$, H$\gamma$, and H$\beta$ (as also highlighted in Fig.~\ref{fig:sedspec}). We note that He\,{\sc i} and [Ne\,{\sc iii}] have also been detected in the spectrum of \galname\ \citep[e.g.][]{Carnall23,Schaerer22}, but we focus on the above transitions, which are used in the primary analysis to infer the physical properties of \galname. 
For each line we derive the fluxes by modeling them with Gaussian profiles, keeping each line flux as a free parameter but tying the redshift and the FWHM broadening together across transitions. This effectively assumes that all the nebular lines originate from the same \hii\ region within the galaxy. In the following, however, we only consider the line ratios of transitions close together in wavelength space to minimize the propagation of any potential offsets in the flux calibration in the derived results. The regions of the spectrum zoomed in on for these particular transitions with the best-fit models included are provided in Appendix B.

\begin{figure}[!t]
\centering
\includegraphics[width=8.8cm]{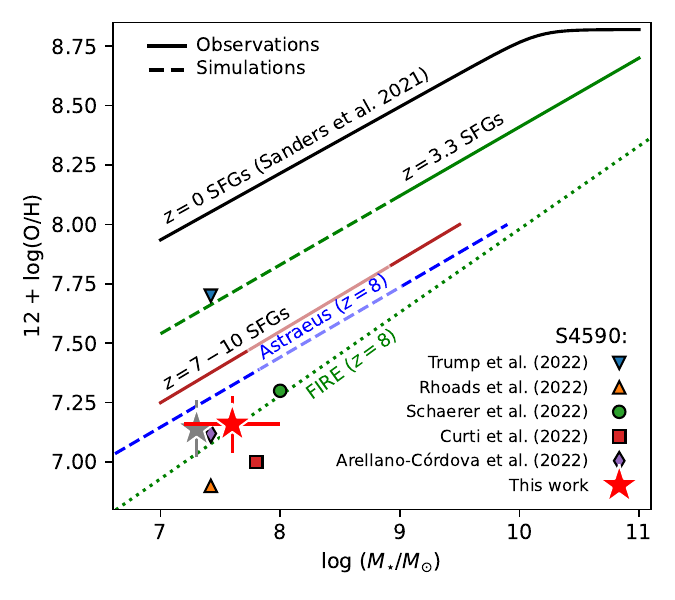}
\caption{Mass-metallicity (MZ) relation of \galname. For comparison, we show the various preexisting MZ estimates for \galname\ from the literature, which highlights the uncertainties from the various methods used. We also show various MZ relations either observed for galaxies at $z\approx 0$ and $z\approx 3.3$ \citep[black and green, respectively;][]{Sanders21} and at $z\approx 7-10$ \citep{Heintz22b} or predicted from the Astraeus \citep{Ucci23} and FIRE \citep[][]{Ma16} simulations, extrapolated to $z\approx 8$. The local, metal-poor galaxy \izw\ is marked by a gray star symbol, showing an MZ relation consistent with that of \galname. Both are consistent with the predictions for the high-$z$ MZ relations.}
\label{fig:massmet}
\end{figure}

From the joint fit, we measure a redshift of $z = 8.4959\pm 0.0003$. The derived Balmer line ratio, ${\rm H}\gamma/{\rm H}\beta = 0.54\pm 0.04$, which is consistent with the theoretically predicted ratio for a Case B recombination scenario \citep{Osterbrock}, further implies a low dust attenuation in \galname\ as also revealed by the best-fit SED model. 
We infer the SFR based on the H$\beta$ flux measurements as 
\begin{equation}
    {\rm SFR}_{\rm H\beta}(M_\odot/{\rm yr}) = 5.5\times 10^{-42} L_{\rm H\beta}({\rm erg/s})\times f_{\rm H\alpha/H\beta} ~ ,
\end{equation}
assuming a \citet{Kroupa01} IMF and the predicted $f_{\rm H\alpha/H\beta} = 2.76$ ratio from the Case B recombination scenario (at $T_e = 20,000$\,K). We derive an intrinsic SFR of ${\rm SFR}_{\rm H\beta} = 2.9\pm 0.9\,M_\odot\,{\rm yr^{-1}}$ based on the photometrically calibrated spectrum and taking into account the magnification factor. The derived SFR and stellar mass are consistent with the star-forming main sequence implied by the parameterization of \citet{Speagle14} when extrapolated to $z=8.5$, in addition to recent results presenting the SFR-$M_\star$ main sequence of galaxies at $z\sim 7-10$ \citep{Heintz22b}. Additionally, the specific SFR (${\rm sSFR} = {\rm SFR}/M_\star$) of $\log{\rm sSFR} = -6.7\pm 0.4\,$yr$^{-1}$ is consistent with recent estimates of galaxies at $z\gtrsim 7$ \citep{Stefanon22,Topping22} and the observed redshift evolution of sSFR($z$) out to $z\sim 8.5$.

To determine the electron temperature $T_e$ and the gas-phase metallicity through the oxygen abundance $12+\log$(O/H) we follow the prescriptions by \cite{Izotov06}. 
We derive $T_e = (2.4\pm 0.4) \times 10^{4}$\,K and $12+\log{\rm (O/H)} = 7.16^{+0.10}_{-0.12}$ (i.e. $\log Z/Z_\odot = -1.53$, assuming $12+\log{\rm (O/H)_\odot} = 8.69$; \citealt{Asplund09}) for all $n_e \lesssim 10^{4}\,$cm$^{-3}$, which is well beyond the inferred electron densities of \hii\ regions in local galaxies \citep{Hunt09} and of galaxies at higher redshifts \citep[$n_e \approx 10^{2}\,$cm$^{-3}$][]{Gillman22}. \citet{Fujimoto23} also found $n_e = 220^{+170}_{-100}\,$cm$^{-3}$ for \galname\ directly from the [\oiii]-$88\mu$m/$\lambda 5007$ line ratio. Our results are generally consistent with previous estimates \citep{Trump22,Schaerer22,Rhoads22,Curti23,ArellaniCordova22,Brinchmann22}, though we note the slightly lower $T_e$ and higher $12+\log$(O/H), which represent the weaker [\oii]\,$\lambda\lambda 3726,3729$ and [\oiii]\,$\lambda 4363$ line transitions detected in our extracted version of the spectrum. 
The oxygen abundance is found to be dominated by O$^{2+}$, with only a small contribution from O$^+$. We further note that the ratio [\oiii]\,$\lambda 5007$/H$\beta = 3.65\pm 0.11$ places \galname\ in the upper region of the Baldwin-Phillips-Terlevich (BPT) diagram \citep{Baldwin81}, suggesting high ionization parameters from harder ionization fields or potential active galactic nucleus contributions. The narrow emission lines seem to exclude the latter, however. Our results thus mainly point to \galname\ being a dense, highly ionized star-forming galaxy. 

We find that the combined mass and metallicity of \galname\ is consistent with those predicted from the Astraeus \citep{Ucci23} and the Feedback in Realistic Environments (FIRE) \citep{Ma16} and simulations, extrapolated to $z\approx 8$, see Fig.~\ref{fig:massmet}, but slightly lower than the recently observed mass-metallicity (MZ) relation of galaxies at $z\approx 7-10$ \citep{Heintz22b}. We note that none of the empirical calibrations of the fundamental SFR-$Z$-$M_\star$ (FMR) plane relation derived for galaxies at $z\approx 0-3$ \citep[e.g.,][]{Curti20,Sanders21,Li22} are able to recover the derived low metallicity of \galname\ \citep[see, e.g.,][]{Curti23}. Our SFR and $M_\star$ measurements for instance predict $12+\log$(O/H) = $7.76$, about 0.6 dex higher than that derived from the strong-line diagnostics. This offset appears to be ubiquitous in high-$z$ galaxies at $z\gtrsim 5$ \citep{Heintz22b}, likely indicating substantial infall of neutral, pristine gas diluting the metallicity at this epoch.
Intriguingly, \galname\ shows a similar pair of mass and metallicity measurements similar to those of the local, extremely metal-poor galaxy I Zwicky 18 (\izw), which is characterized by an oxygen abundance of $12+\log{\rm (O/H)} = 7.14$ and $M_\star = 2\times 10^{7}\,M_\odot$ \citep{Madden13}. 

Determining the gas-phase metallicity from the description by \citet[][their eq. 2]{Jones20} relying on the FIR [\oiii]-$88\mu$m transition yields $12+\log{\rm (O/H)} = 7.82$. This method thus overpredicts the metallicity by 0.65 dex as compared to the results based on the direct $T_e$ method derived from the redshifted optical nebular and auroral emission lines here. Comparing this metallicity to other strong-line diagnostics defined for local ``analogs" of high-redshift galaxies \citep{Bian18} yields offsets of $-0.2$, $0.4$, and $-0.3$ dex for the R23, O32, and O3H$\beta$ calibrations, respectively. The most accurate calibration for this particular galaxy seems to be the O32 calibration from the Sloan Digital Sky Survey reference sample in \citet{Bian18}, similar to the O32 diagnostics derived by \citet{Curti20}, which seems to be consistent with the direct metallicity within 0.1 dex \citep[as is also evident from the galaxy at $z= 9.5$ presented by][]{Heintz22b}.  

\subsection{The total dynamical mass}

Assuming that [\cii] traces the ISM of \galname\, we can further derive the total dynamical mass $M_{\rm dyn}$ of this galaxy.  We caution, however, that the spectroscopic blueshift and the $\approx 2\sigma$ spatial offset of [\cii] relative to the UV component and [\oiii] emission \citep{Fujimoto23} indicate that the origin of the [\cii] emission is not fully co-spatial with the most intense, star-forming region of the galaxy. Such extended, blueshifted [\cii] emission has been observed in other high-$z$ galaxies, and may be hinting at an offset, abundant gas reservoir, a shock-heated outflowing gas, or minor mergers \citep[e.g.,][]{Maiolino15,Kohandel19,Arata20,Akins22,Katz22b}. The latter scenario would typically produce much broader ($>1000$\,km\,s$^{-1}$) [\cii] line profiles \citep{Appleton13}, whereas the observed narrow-line feature is consistent with the low stellar mass of \galname. We also caution that we are not able to firmly exclude the presence of a merging system due to the limited spatial resolution of our ALMA observations \citep{Rizzo22}. However, given that the [\cii] line width and luminosity are consistent with predictions based on the stellar mass and SFR of \galname, we assume in the following that the [\cii] emission is physically associated with the ISM of the galaxy.

We infer $M_{\rm dyn}$ based on the measured FWHM of the [\cii] emission line corrected by the instrumental resolution, FWHM$_{\rm [CII]} = 114\pm 35$\,km\,s$^{-1}$, as 
\begin{equation} \label{eq:mdyn}
    M_{\rm dyn} = 1.16\times 10^{5} v^2_{\rm circ} D_{\rm [CII]}
\end{equation}
following, e.g., \citet[][]{Wang13} and \citet{Willott15}. Here we approximate the circular velocity to $v_{\rm circ} \approx 0.52\times {\rm FWHM}_{\rm [CII]}$, appropriate for velocities dominated by nonordered rotation \citep{Decarli18,Neeleman21}. \citet{Fujimoto23} measured a circularized [\cii] effective radius of $0\farcs69\pm 0\farcs42$ in the image plane, which corresponds to $0\farcs 23 \pm 0\farcs14$ in the source plane or to a physical size of $r_{\rm e,[CII]} = 1.1\pm 0.7$\,kpc at $z=8.496$. Such extended [\cii] halos reaching beyond the UV components of the galaxies are typically observed at $z\gtrsim 6$, with $r_{\rm e,[CII]} \approx 2-3 \times r_{\rm e,UV}$ \citep{Smit18,Fujimoto19,Matthee19,Carniani20,Fudamoto22,Akins22}, though \galname\ presents one of the most notable cases with $r_{\rm e,[CII]} \approx 10 \times r_{\rm e,UV}$. Assuming the ``disk" diameter is $D_{\rm [CII]} = 2\times r_{\rm e,[CII]}$, we infer $D_{\rm [CII]} = 2.2\pm 1.4$\,kpc. This yields a dynamical mass of $M_{\rm dyn} = (9.0\pm 4.5)\times 10^{8}\,M_\odot$. We note that if we assume ordered, rotation-dominated motion and an average inclination angle of 45$^{\circ}$, the dynamical mass increases by $\approx 50\%$. However, since there is no strong evidence of ordered motion in \galname, we adopt the above-derived $M_{\rm dyn}$. 

\section{The gas, dust, and metal content} \label{sec:disc}

\subsection{Disentangling the baryonic matter budget}

Based on the assumption that the dynamical mass is dominated by the baryonic matter content within the [\cii]-emitting region \citep[which is confirmed to be the case for dusty, massive galaxies at $z\sim 4-5$;][]{Rizzo20,Rizzo21}, the total dynamical mass of \galname\ subtracted by its stellar mass should represent the total gas content of this galaxy (in molecular and/or neutral, atomic form); $M_{\rm gas} = M_{\rm dyn}-M_\star$. For \galname, the stellar mass derived from the SED-fitting only constitutes $\approx 1-2\%$ of the total dynamical mass, such that $M_{\rm gas} \approx M_{\rm dyn}$. However, using the more conservative estimate of $M_\star \approx 10^{8}\,M_\odot$ derived from more complex SFHs or spatially resolved analyses \citep{Tacchella22,GimenezArteaga23}, we estimate a total gas mass of $M_{\rm gas} = (8.0\pm 4.0)\times 10^{8}\,M_\odot$. This yields a gas mass excess and gas fraction of $\log(M_{\rm gas}/M_\star) = 0.9-1.7$ and $f_{\rm gas} = M_{\rm gas}/(M_{\rm gas} + M_\star) \gtrsim 90\%$. This gas mass excess is substantial but not too surprising given the overall increase of $M_{\rm gas}/M_\star$ with increasing redshift \citep{Geach11,Carilli13,Scoville17,Tacconi13,Tacconi18,Heintz21,Heintz22} and decreasing metallicity \citep{Catinella18,Stark21,Heintz21}. 

\begin{figure}[!t]
\centering
\includegraphics[width=8.8cm]{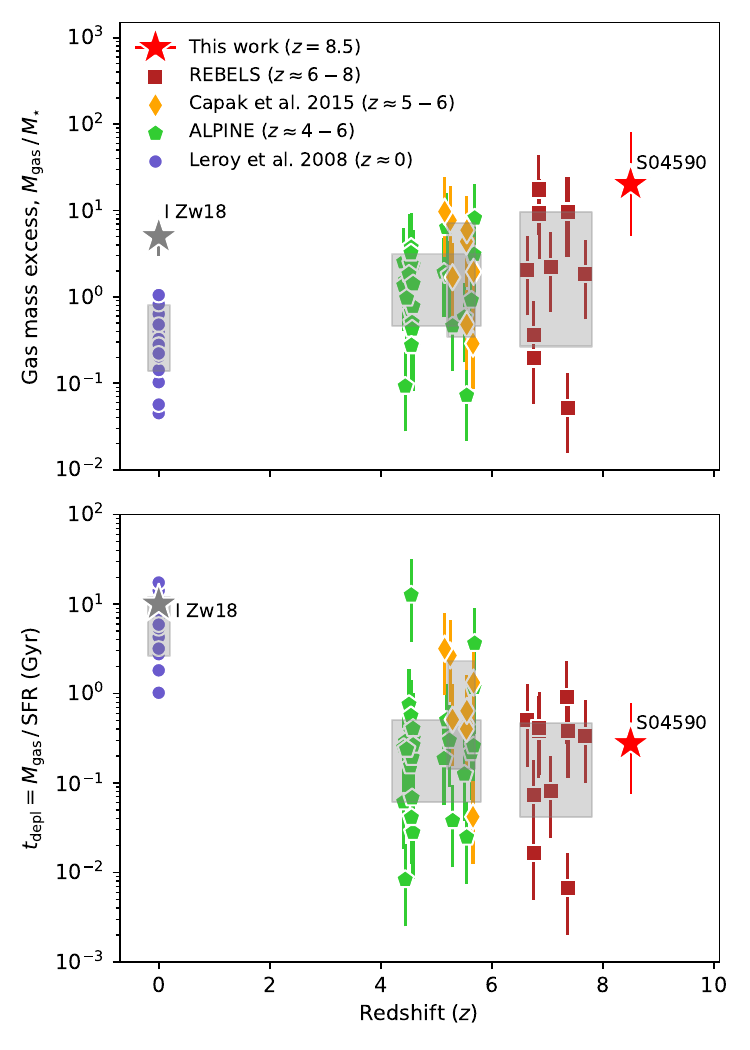}
\caption{(Top panel): Total gas mass excess, $M_{\rm gas}/M_\star$, as a function of redshift. For the local galaxy sample at $z\sim 0$ \citep{Leroy08}, $M_{\rm gas} = M_{\rm HI} + M_{\rm H_2}$. For the high-$z$ galaxy sample compilation (see main text for references), we assume that the total gas mass can be estimated as $M_{\rm gas} = M_{\rm dyn,tot} - M_\star$, with $M_{\rm dyn,tot}$ estimated from the [\cii] line width and spatial extent. Gray boxes indicate the redshift distributions and standard deviations for each sample. Overall, $M_{\rm gas}/M_\star$ increases with redshift by a factor of $\approx 10$ from $z\sim 0$ to $z\sim 7$, and $\log(M_{\rm gas}/M_\star) = 0.9-1.7$ for \galname. 
(Bottom panel): Total gas depletion time, $t_{\rm depl} = M_{\rm gas}/{\rm SFR}$, as a function of redshift; $t_{\rm depl}$ is observed to decrease by a factor of $\approx 10$ on average from $z\sim 0$ to $z\sim 7$.
}
\label{fig:mhiz}
\end{figure}

\begin{figure}[!t]
\centering
\includegraphics[width=8.8cm]{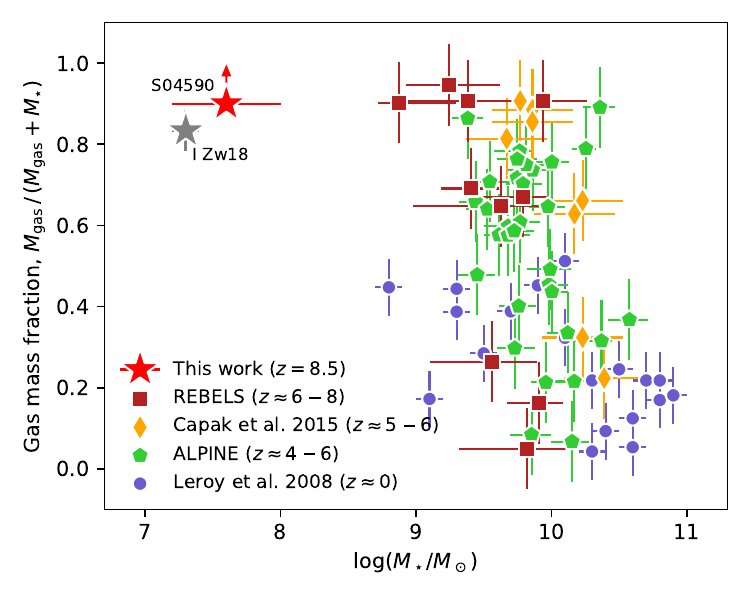}
\caption{Gas mass fraction, $f_{\rm gas} = M_{\rm gas}/(M_{\rm gas}+M_\star)$, as a function of stellar mass. The symbol notation follows Fig.~\ref{fig:mhiz}. We observe an overall anticorrelation of $f_{\rm gas}$ with stellar mass for both the low- and high-$z$ galaxy samples, though with an offset due to the redshift evolution of $M_{\rm gas}/M_\star$. \galname\ and \izw\ are both clear outliers, with low stellar masses $M_\star = 1-2\times 10^{7}\,M_\odot$ and high gas fractions $f_{\rm gas}\gtrsim 90\%$, but seem to follow the overall trend.
}
\label{fig:fgas}
\end{figure}

In Fig.~\ref{fig:mhiz} we show the total gas mass excess and gas depletion of \galname\ together with a set of literature data as a function of redshift. We include the local sample of galaxies at $z\sim 0$ by \citet{Leroy08}, for which we estimate $M_{\rm gas} = M_{\rm HI} + M_{\rm H_2}$. At high $z$ ($z\gtrsim 4$), we compile galaxy samples for which the dynamical mass and the [\cii]-emitting size can be inferred, which include the The ALMA-ALPINE survey \citep[$z\sim 4-6$;][]{LeFevre20,Bethermin20,Faisst20}, the sample from \citet[][$z\sim 5-6$]{Capak15}, and the ALMA-REBELS survey \citep[$z\sim 6-8$;][]{Bouwens22}. These all target ``regular'' star-forming galaxies at the respective survey redshifts. To estimate $M_{\rm gas}$ for the high-$z$ sample compilation, we again assume $M_{\rm gas} = M_{\rm dyn}-M_\star$ with $M_{\rm dyn}$ estimated using Eq.~\ref{eq:mdyn} and $M_\star$ as provided in the catalogs. If they are not reported, we assume inclination angles of $45^\circ$ for the [\cii]-emitting galaxy compilation. We observe a strong increase in the gas mass excess as a function of redshift, with averages of $\log(M_{\rm gas}/M_\star) = -0.3$ at $z\approx 0$ and $\log(M_{\rm gas}/M_\star) = 0.7$ at $z\approx 6-8$, respectively. Similarly, we compute the gas depletion times for the compiled galaxy samples and find averages of $t_{\rm depl} = 7\,$\,Gyr at $z\approx 0$, declining to $t_{\rm depl} = 0.2$\,Gyr at $z\gtrsim 4$. In both cases, \galname\ is a clear outlier owing to its large inferred gas mass. However, this can naturally be explained by its low metallicity and stellar mass, for which a large gas fraction is expected, as shown in Fig.~\ref{fig:fgas}, which are consistent with the properties of \izw.

We now attempt to further disentangle the baryonic mass components of \galname\ by estimating the contributions from the cold molecular and neutral atomic gas phases to the total gas mass. Recently, \citet{Zanella18} calibrated the [\cii] luminosity to the total molecular gas mass ($M_{\rm mol}$) for a set of galaxies at lower redshifts ($z\sim 0-2$). They found a conversion factor of $\alpha_{\rm [CII]} = $ [\cii]-to-$M_{\rm mol} = 30\, M_{\odot}/L_\odot$ with a scatter of 0.3 dex \citep[see also][]{Madden20}, independent of the metallicity. Simulations of galaxies at $z\sim 6$ suggest a slightly lower but consistent average conversion of $\alpha_{\rm [CII]} = 18\, M_{\odot}/L_\odot$ \citep{Vizgan22a}. Applying these calibrations to $L_{\rm [CII]}$ yields $M_{\rm mol} = (3.0-5.0)\times 10^{8}\,M_\odot$ (depending on the conversion used), with a combined uncertainty of $\approx 40\%$ taking into account the statistical uncertainty and scatter in the calibration. The molecular gas content thus comprises $\approx 30-50\%$ of $M_{\rm gas}$ by mass. 

However, [\cii] can also be used as a proxy for the neutral atomic hydrogen \hi\ gas mass \citep{Heintz21,Heintz22,Vizgan22b} since it has been shown observationally to predominantly originate from the neutral gas phase in the Milky Way and nearby
galaxies \citep{Madden93,Madden97,Pineda14,Croxall17}, with supporting evidence from high-redshift galaxy observations \citep[e.g.,][]{Novak19,Meyer22} and simulations \citep{Katz17,RamosPadilla22} as well. We adopt the [\cii]-to-\hi\ calibration from \cite{Heintz21}, prescribed as
\begin{equation}
\begin{split}
    \log \beta_{\rm [CII]} = \log M_{\rm HI} / L_{\rm [CII]} = (-0.87\pm 0.09) \times \\ \log(Z/Z_{\odot}) + (1.48\pm 0.12),
    \label{eq:ciitohi}
\end{split}
\end{equation}
see also \citet{Heintz22}. In this previous work, the $M_{\rm HI}$ of [\cii]-emitting galaxies at $z\sim 6-8$ from the REBELS survey \citep[e.g.,][]{Bouwens22} could only be inferred based on uncertain metallicity estimates assuming an SFR-$Z$-$M_\star$ fundamental plane relation \citep[e.g.,][]{Curti20}. With the accurate gas-phase metallicity from the NIRSpec analysis performed here, we obtain the first robust high-$z$ $M_{\rm HI}$ estimate of $M_{\rm HI}\approx  10^{10}\,M_\odot$. This is $\approx 10\times$ more massive than the inferred dynamical mass. This can either be explained by {\it i)} the [\cii]-to-\hi\ relation being less robust at these high redshifts or low metallicities; {\it ii)} the existence of a much larger \hi\ gas component situated at larger radii than assumed for the [\cii]-emitting disk, which then causes the total dynamical mass of the system to be underestimated; or {\it iii)} an increased excitation of $L_{\rm [CII]}$. 
However, the derived ratio $L_{\rm [CII]}/{\rm SFR} \approx 10^{7}\,L_\odot / (M_\odot\,{\rm yr^{-1}})$ \citep[see also][]{Fujimoto23} appears consistent with the ubiquitously observed SFR$-L_{\rm [CII]}$ relation \citep[e.g.,][]{DeLooze14,Schaerer20}. 
We also note that \izw\ again has remarkably comparable properties in terms of the metallicity ($12+\log$(O/H) = 7.14), and with an estimated \hi\ gas mass of $M_{\rm HI} = 10^{8}\,M_\odot$ and [\cii] luminosity $L_{\rm [CII]} = 1.1\times 10^{5}\,L_\odot$ \citep[e.g.,][]{RemyRuyer14,Cormier15}, yielding an equally high [\cii]-to-\hi\ ratio of $\log \beta_{\rm [CII]} = \log (M_{\rm HI}/L_{\rm [CII]}) = 2.95\,M_\odot/L_\odot$. In any case, since $M_\star$ and $M_{\rm mol}$ only constitute $2-10\%$ and $30-50\%$ of the total dynamical mass of \galname, respectively, we surmise that the largest baryonic mass component is likely in the form of neutral atomic \hi\ gas.

\subsection{The dust and metal yield}

With the first high-$z$ combination of the gas-phase metallicity and the total gas mass, we can now infer the ISM {\em metal} mass, $M_{\rm Z,ISM}$, of \galname\ at an unprecedented redshift of $z = 8.5$ \citep[previously only possible up to $z\approx 3$;][]{Sanders23}.
We determine this as
\begin{equation}
    M_{\rm Z,ISM} = M_{\rm gas} \times 10^{12+\log{\rm (O/H)-8.69}} \times Z_\odot, 
    \label{eq:mmets}
\end{equation}
where $Z_\odot = 0.0134$ is the solar metallicity by mass. This yields $M_{\rm Z,ISM} = (3.2\pm 1.6)\times 10^{5}\,M_\odot$. The expected total Type II supernova metal production estimated by \citet{Peeples14}, assuming the prescribed SFHs by \citet{Leitner12} and a nucleosynthetic yield of $y=0.033$, is $M_{\rm Z,SN} = (7.6\pm 1.9)\times 10^{5}\,M_\odot$, marginally consistent with our estimate. Assuming a briefer star-formation history following \citet{Sanders23}, which is likely more appropriate for \galname\ given its high redshift, we obtain consistent results.
This indicates that the majority of the produced metals from core-collapse supernovae in \galname\ are retained in the ISM. This is in contrast with star-forming galaxies at $z\sim 2-3$, which only retain $\approx 20\%$ of the produced metals in the ISM \citep[][]{Sanders23}, and even more so with local galaxies, which have most of their metals captured in stars. Further, this puts strong constraints on the possibility of significant outflows in \galname, as most of the produced metals appear to still be confined to the ISM. 

\begin{figure}[!t]
\centering
\includegraphics[width=9cm]{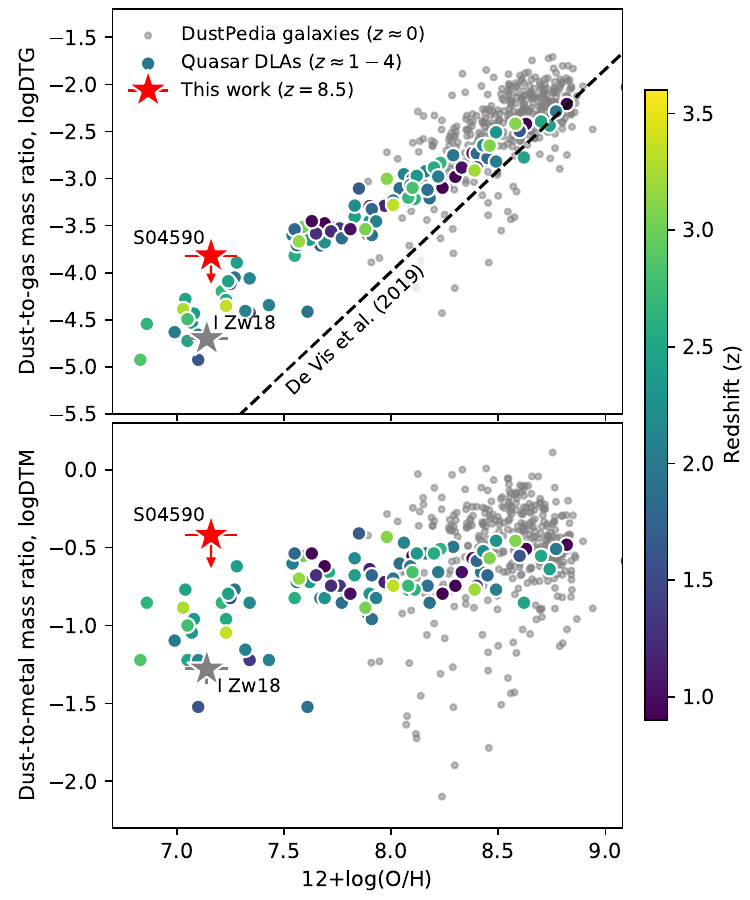}
\caption{DTG (top) and DTM (bottom) mass ratios as a function of the gas-phase metallicity, $12+\log$(O/H). The red and gray star symbols highlight \galname\ and \izw, respectively. The gray dots represent the local $z\approx 0$ {\sc DustPedia} galaxy sample from \citet{DeVis19}, with their best-fit DTG-metallicity relation shown by the dashed line. The colored filled circles show line-of-sight measurements through quasar absorbers at $z\approx 1-4$ \citep{Peroux20}, color-coded as a function of redshift. Both \galname\ and \izw\ are among the most dust-poor galaxies known, but are consistent with the overall evolution with metallicity.   
}
\label{fig:dtgdtm}
\end{figure}

In combination with the constraints placed on the dust mass of $M_{\rm d} < 1.2\times 10^{5}\,M_\odot$ utilizing the FIR continuum coverage provided by ALMA from \citet{Fujimoto23}, we further determine upper bounds on the mass ratios of the dust-to-gas (DTG, $M_{\rm d}/M_{\rm gas}$) and dust-to-metal (DTM, $M_{\rm d}/M_{\rm Z}$) content of \galname. We find $\log$DTG $ < -3.8$ and $\log $DTM$ < -0.5$, respectively, as also shown in Fig.~\ref{fig:dtgdtm}. These DTG and DTM mass ratios are substantially lower than the Milky Way average of ${\rm DTG} \approx 10^{-2}$ and ${\rm DTM} \approx 0.5$. However, this is likely due to metallicity dependence of the DTG and DTM ratios predicted from simulations \citep{Popping17,Vijayan19,Hou19,Li19,Triani20,Graziani20} and observed in local galaxies \citep{RemyRuyer14,DeVis19} and high-$z$ sightlines in absorption \citep{DeCia16,Wiseman17,Peroux20,Popping22}. Comparing our measurements to those from the {\sc DustPedia}\footnote{\url{http://dustpedia.astro.noa.gr/}} sample of local galaxies at $z\approx 0$ \citep{DeVis19} and to absorption-derived ratios from quasar absorption line systems (DLAs) at $z\approx 1-4$ \citep{Peroux20,Popping22} in Fig.~\ref{fig:dtgdtm}, we find that they are indeed consistent with the expected evolution with metallicity that appears universal across redshifts.  
Moreover, these results are again in good agreement with the mass ratios derived for \izw, with an estimated dust mass of $M_{\rm dust} \approx 2\times 10^{3}\,M_\odot$ \citep{Cannon02}, yielding a DTG ratio of $2\times 10^{-5}$ and a DTM ratio$\approx 5\%$. These results highlight the strong deviation of the DTG and DTM mass ratios from the Milky Way average at high redshifts and low metallicities.

\section{Conclusions} \label{sec:conc}

We have presented a comprehensive characterization of the gas, metal, and stellar content of a galaxy at $z_{\rm spec} = 8.496$ (dubbed \galname), providing the most detailed view to date of a galaxy well into the epoch of reionization. 
These results were enabled by the first joint analysis of the exquisite photometric and spectroscopic observations obtained with the NIRCam, NIRISS, and NIRSpec instruments on board \jwst\ from the early-release science observations \citep{Pontoppidan22} and follow-up ALMA observations of the [\cii]$-158\mu$m line transition (through the DDT program: 2021.A.00022.S, PI: S. Fujimoto). We derived the physical properties of \galname\ by modeling the SED of the galaxy and through the strong nebular and auroral emission lines detected in the spectrum. We found that it is a low-mass ($M_\star = 10^{7.2}-10^{8.0}\,M_\odot$), low-metallicity (12+log(O/H) = $7.16^{+0.10}_{-0.12}$) galaxy, which appears to be overall consistent with predictions for the MZ relation and the SFR$-M_\star$ main sequence at the observed redshift and previous literature estimates, but not with those for the SFR-$M_\star$-$Z$ relation at lower redshifts.

With the simultaneous detection of [\cii], we further constrained the dynamics and gas content of the ISM in \galname. First, we determined a total dynamical mass of $M_{\rm dyn} = (9.0\pm 4.5)\times 10^{8}\,M_\odot$ based on the [\cii] line width. Subtracting the inferred stellar mass, contributing only $\approx 2-10\%$ to the total dynamical mass, we determined a total gas mass of $M_{\rm gas} = (8.0\pm 4.0)\times 10^{8}\,M_\odot$. \galname\ was thus observed to have a gas mass excess and gas fraction of $\log(M_{\rm gas}/M_\star) = 0.9-1.7$ and $f_{\rm gas} = M_{\rm gas}/(M_{\rm gas} + M_\star) \gtrsim 90\%$, respectively. While these results indicate that \galname\ contains one of the most extreme gas fractions derived for high-$z$ galaxies to date, we demonstrated that they are in perfect agreement with the inferred low stellar mass and gas-phase metallicity. We then attempted to disentangle the various gas-phase components through established conversion factors relying on the [\cii] line luminosity, for which we inferred $M_{\rm mol} = (3.0-5.0)\times 10^{8}\,M_\odot$ (i.e. $30-50\%$ of $M_{\rm gas}$). We surmised that the largest contributing factor to the gas mass and the overall baryonic matter budget is thus in the form of neutral, atomic \hi, as supported by recent [\cii]-to-\hi\ scaling relations. 

With the joint estimates of the gas-phase metallicity and the gas mass, we further estimated the total mass of metals in the ISM of \galname, finding $M_{\rm Z,ISM} = (3.2\pm 1.6)\times 10^{\rm 5}\,M_\odot$. This is consistent with the expected Type II core-collapse supernova metal yield for the inferred stellar mass build-up. With the independent estimate of the dust mass from the FIR continuum, $M_{\rm d} < 1.2\times 10^{5}\,M_\odot$ \citep{Fujimoto23}, we constrained the mass ratios of the DTG and DTM abundance finding $\log$DTG $ < -3.8$ and $\log$DTM $ < -0.5$. This places \galname\ as one of the most dust-poor galaxies known to date relative to its inferred gas and metal mass and demonstrates a strong deviation from the typical Milky Way averages in the high-redshift and low-metallicity regimes.

The overall physical properties of \galname\ were also found to show remarkable similarity to those of the local, metal-poor galaxy I Zw 18 in terms of the MZ relation, gas fraction, and overall gas, dust, and metal abundance. Generally, this highlights the overall resemblance of the emission line properties of these high-$z$ metal-poor galaxies to local analogs \citep[as also showcased by, e.g.,][]{Schaerer22,ArellaniCordova22,Katz23}, and suggests that galaxy properties and their overall evolution are governed by the same fundamental physical processes.  
With the advent of \jwst\ and in synergy with ALMA, we are now in a position to characterize the stellar, gas, and metal content and constrain how these processes govern the formation and evolution of the first generation of galaxies, as demonstrated by this work. 


\section*{Acknowledgements}

We would like to thank the referee for providing a constructive, greatly improving the presentations of the results in this paper. Further, we would like to thank Fengwu Sun for providing insightful comments on an early draft of this project.
K.E.H. acknowledges support from the Carlsberg Foundation Reintegration Fellowship Grant CF21-0103 and VILLUM FONDEN through the Villum Experiment Programme.
The Cosmic Dawn Center (DAWN) is funded by the Danish National Research Foundation under grant No. 140. 
T.R.G. acknowledges support from the Carlsberg Foundation (grant no CF20-0534).
K.K acknowledges the support by JSPS KAKENHI Grant Number 17H06130 and the NAOJ ALMA Scientific Research Grant Number 2017-06B.

This work is based in part on observations made with the NASA/ESA/CSA James Webb Space Telescope. The data were obtained from the Mikulski Archive for Space Telescopes (MAST) at the Space Telescope Science Institute, which is operated by the Association of Universities for Research in Astronomy, Inc., under NASA contract NAS 5-03127 for JWST. These observations were obtained as part of the Early Release Observations (ERO) covering the SMACS J0723.3–7327 galaxy cluster. 
This paper makes use of the following ALMA data: ADS/JAO.ALMA\#2021.A.00022.S. ALMA is a partnership of ESO (representing its member states), NSF (USA) and NINS (Japan), together with NRC (Canada), MOST and ASIAA (Taiwan), and KASI (Republic of Korea), in cooperation with the Republic of Chile. The Joint ALMA Observatory is operated by ESO, AUI/NRAO and NAOJ.

\section*{Data availability statement} 

Source codes for the figures and tables presented in this manuscript are available from the corresponding author upon reasonable request. The reduced and calibrated JWST spectroscopic and imaging data are made available on the dedicated repository at \url{https://github.com/keheintz/S04590}.  
The overall JWST data products are available via the Mikulski Archive for Space Telescopes (\url{https://mast.stsci.edu}). The specific observations analyzed in this work can be accessed via \dataset[DOI:10.17909/67ft-nb86]{https://archive.stsci.edu/doi/resolve/resolve.html?doi=10.17909/67ft-nb86}. \\

\software{This work made use of and acknowledge the following software: {\tt NumPy} \citep{Numpy}, {\tt Matplotlib} \citep{matplotlib}, \texttt{grizli} \citep{Brammer19}, {\tt Astrodrizzle} \citep{AstroDrizzle}, CASA \citep[v6.4.1.12;][]{casa2022}, NIRSpec analyis tools \citep[v0.3;][]{msaexp}, and \textsc{Bagpipes}  \citep{bagpipes}. }


\bibliography{ref}
\bibliographystyle{aasjournal}


\begin{appendix}

\section{Best-fit SED model}

\begin{figure*}[!h]
\centering
\includegraphics[width=15cm]{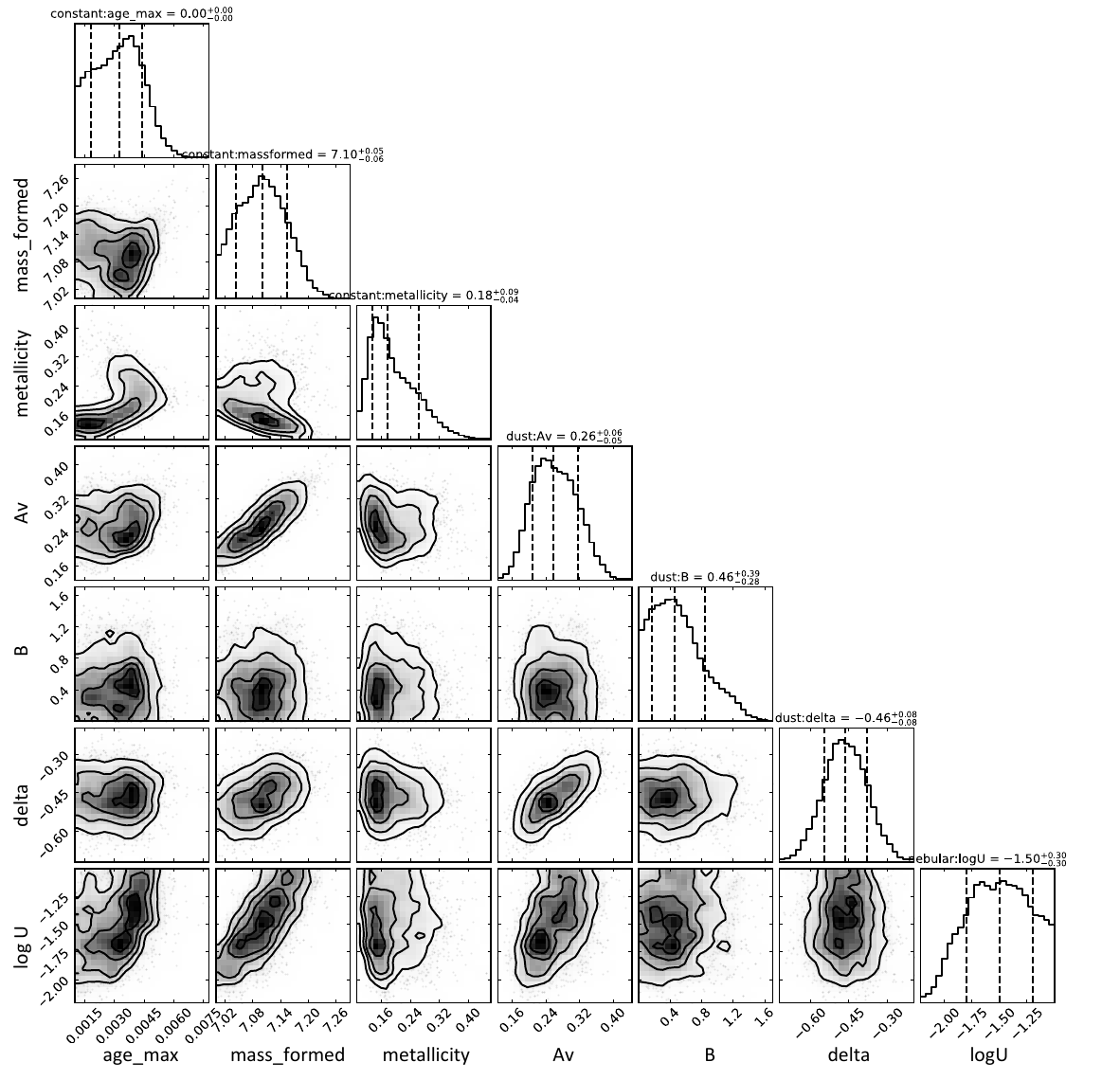}
\caption{Cornerplot of SED posterior distributions from {\sc Bagpipes}.}
\label{fig:sedcorner}
\end{figure*}

As briefly described in the main text, we model the SED of \galname\ using {\sc Bagpipes} based on the measured \jwst/NIRCam+NIRISS photometry. Specifically, we assume a \cite{Salim18} dust model, allowing the slope ($\delta$) to vary within $-0.9<\delta<0.1$, and the 2175Å bump strength ($B$) to vary within $0<B<3$. We assume Gaussian priors on the various dust parameters. The visual extinction, which we allow to vary in a linear grid between 0 and 2, has the prior centered at 0.1, with a sigma of 0.5, which is the same prior as that for $B$. For $\delta$, we impose a Gaussian prior centered at $-0.5$, with a $0.1\sigma$. We also allow the metallicity to vary within $0<Z<1$, with a prior centered at 0.2 with a $0.1\sigma$. We assume a \cite{Kroupa01} IMF. For the constant SFH model, we set the maximum age in a grid from 1~Myr to 10~Myr. Finally, we set the lifetime of birth clouds to 10~Myr. Increasing the maximum age allowed in the prior to 100~Myr we still find the code to infer a mass-weighted age $<10$\,Myr ($2.2\pm 0.8$\,Myr), and the stellar mass remains consistent with the result presented in the main text. We also note that other literature studies have found consistently young ages, using a similarly constant SFH \citep[][]{Carnall23} or a non-parametric SFH \citep[][]{Tacchella22}.   
The resulting posterior distributions of the best fit are shown in Fig.~\ref{fig:sedcorner}, and summarized in Table~\ref{tab:galprops}. 

\section{Line flux estimates}

\begin{figure}[!h]
\centering
\includegraphics[width=9cm]{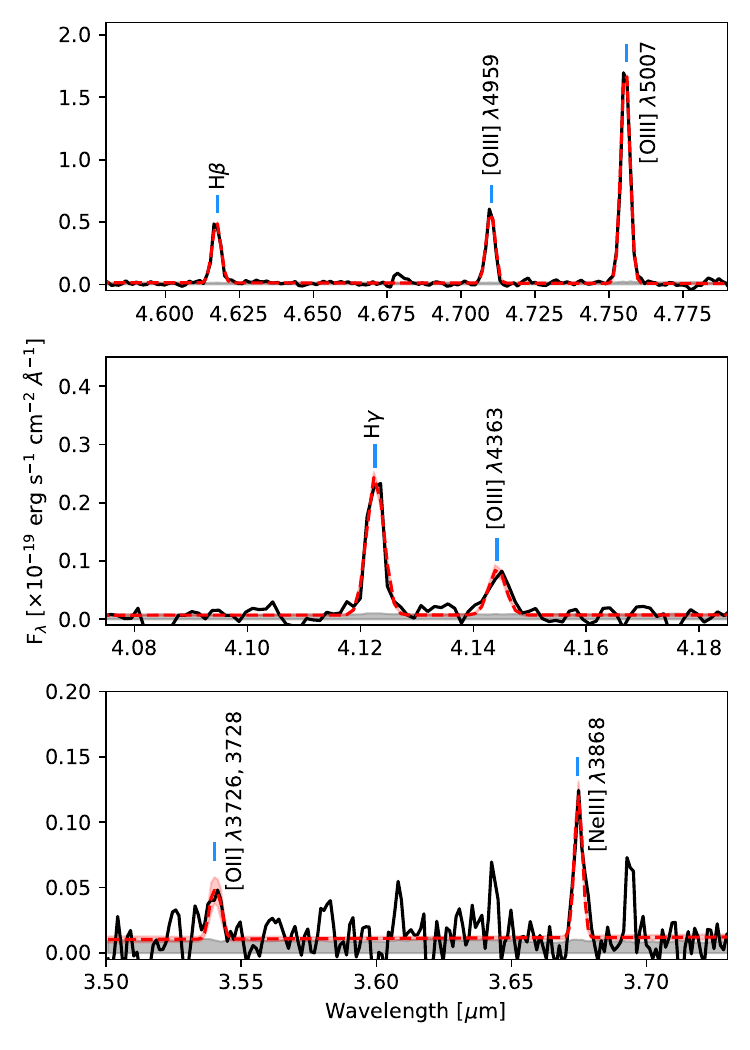}
\caption{Zoomed-in views of the regions of the spectrum that include the main nebular and auroral emission line transitions examined in this work. The black line shows the flux density, with the gray region representing the noise level. The best-fit continuum and Gaussian profiles to each line transition are shown by the dashed red line. The [\oii]\,$\lambda\lambda 3726,3728$ doublet is not resolved at this spectral resolution.}
\label{fig:linefit}
\end{figure}

In Fig.~\ref{fig:linefit} we show regions of our extracted NIRSpec spectrum, zoomed in on the relevant nebular and auroral Balmer and oxygen line transitions. Overplotted are also the best-fit continuum models and Gaussian profiles used to extract the line fluxes that the main analysis is based on. 

\end{appendix}

\end{document}